\documentstyle[12pt]{article}

\parskip=\medskipamount
\begin{document}

\begin{titlepage}

\begin{center}
{ \large \bf Can pions created in high-energy heavy-ion collisions produce
a Centauro-type effect?  }\\
\vspace{2.cm}
{ \normalsize \bf M. Martinis, \footnote { e-mail address:
martinis@thphys.irb.hr}
 V. Mikuta-Martinis, \footnote {e-mail address: mikutama@thphys.irb.hr}
and J. \v Crnugelj \footnote
{e-mail address:crnugelj@thphys.irb.hr }} \\
\vspace{0.5cm}
Department of Physics, \\
Theory Division, \\
Rudjer Bo\v skovi\' c Institute, P.O.B. 1016, \\
41001 Zagreb,CROATIA \\
\end{center}
{ \large \bf Abstract}
\vspace{0.5cm}
\baselineskip=20pt

We study a Centauro-type phenomenon in high-energy heavy-ion collisions
by assuming that pions are produced semiclassically both directly
and in pairs through the isovector channel.

The leading-particle effect and the factorization property of the
scattering amplitude in the impact-parameter space are used to define
the classical pion field. By analyzing the joint probability function
$P_{II_{3}}(n_{0},n_{ \_})$ for producing $n_{0}$ neutral and $n_{-}$
negative pions from a definite isospin state $II_{3}$ of the incoming
leading-particle system we show that only direct production of pions
without isovector pairs favors Centauro-type behavior. The presence of
isovector pairs seems to destroy the effect. Our conclusion is
supported through the calculation of two pion correlation parameters,
$f_{2}^{0-}$ and $f_{2}^{00}$, and the average number of neutral pions
$( \langle n_{0} \rangle_{n_{ \_}})$ as a function of negative pions
$(n_{ \_})$ produced.

\vspace{2cm}

\end{titlepage}
\newpage
\baselineskip=24pt

In recent years several cosmic-ray experiments [1] have reported evidence
for the existence of Centauro events characterized by an anomalously large
number of charged pions in comparison with the number of neutral pions,
indicating that there should exist a very strong long-range correlation
between two types of the pions. Such long-range correlations are possible
if pions are produced coherently and constrained by the global conservation of
isospin [2--7].

\setcounter{page}{2}
Although the actual dynamical mechanism of the
production of a classical pion field in the course
of a high-energy collisions is not known, there exist a
number of interesting theoretical speculations [8--13] that localized
regions of misaligned vacuum might occur in ultrahigh-energy hadronic
and heavy-ion collisions. These regions become coherent sources of a
classical pion field. In early models [2,3], however, the coherent
production of pions is taken for granted and is
considered to be a dominant mechanism.

These models  also predict strong negative correlations between the number
of neutral and charged pions. In fact, the exact conservation of isospin
in a pion uncorrelated jet model is known [4,5] to
give the same pattern of neutral/charged fluctuations
as observed in Centauro events. This strong negative
neutral-charged correlation is believed to be a
general property of the direct pion emission in
which the cluster formation ( or the short-range correlation between pions )
is not taken into account [6,7,14].

A similar conclusion may be drawn when a proper
multipion symmetrization combined with the isospin singlet channel
is considered [15].

In this paper we consider the leading-particle effect as a
possible source of a classical pion field [16]. Pions are
assumed to be produced from a definite isospin state of the
incoming leading-particle system both directly and in
isovector pairs.

Coherently emitted isovector clusters decay subsequently into
pions outside the region of interaction. We discuss the
limiting behavior of the probability distribution of
neutral pions, the neutral - to - charged ratio and the
corresponding two-pion correlation function $f_{2}^{ab},$
as well as the variation of the average number of neutral pions
$( \langle n_{0} \rangle_{n_{ \_}})$ as a function of the
number of negative pions $(n_{ \_})$ produced.

We show that the Centauro effect is strongly suppressed if pions are
produced in isovector pairs.
At high energies most of the pions are
produced in the central region. To isolate the central production,
we adopt high-energy longitudinally dominated kinematics, with
two leading particles retaining a large fraction of their incident
momenta. The energy available for the hadron production is
\begin{equation}
E_{had} = \frac{1}{2} \sqrt{s} - E_{leading}.
\end{equation}
It is related to the inelasticity $K$ of the collisions by
\begin{equation}
E_{had} = \frac{1}{2} \sqrt{s}K.
\end{equation}
Note that at fixed total c.m. energy $ \sqrt{s},$ the hadronic energy
$E_{had}$ varies from event to event.

With a set of independent variables $s, \; \{
\vec{q}_{iT},y_{i} \} \equiv q_{i}, \; \; i = 1,2, \ldots n,$
the n-pion contribution to the $s$-channel unitarity becomes an
integral over the relative impact parameter  $b$ of the two
incident leading particles:
\begin{equation}
F_{n}(s) = \frac{1}{4s} \int d^{2}b \prod_{i=1}^{n}dq_{i}
\mid T_{n}(s, \vec{b};1 \ldots n) \mid^{2},
\end{equation}
where $ dq = d^{2}q_{T}dy/2(2 \pi)^{3}.$ The normalization is
such that
\begin{eqnarray}
F_{n}(s) & = & s \sigma_{n}(s), \nonumber \\
\sigma_{inel}(s) & = & \sum_{n=1}^{ \infty} \sigma_{n}(s).
\end{eqnarray}
At high energies the matter distributions of the two-leading-particle
system are Lorentz-contracted disks in the c.m. system.
We assume that, in the early stages of the collision, a highly excited
localized system occurs that relaxes through the coherent emission of pions
[17]. According to [18], this type of coherent emission of
pions should saturate close to the threshold since,
once it starts, the resulting pion emission, in the
absence of a " resonance cavity ", prevents further
buildup of pion fields.
The basic assumption of the independent pion-emission model,
neglecting the isospin for a moment, is the factorization of the
scattering amplitude $ T_{n}( s, \vec{b}; 1 \ldots n )$ in the
$b$ space:
\begin{equation}
T_{n}( s, \vec{b}; 1 \ldots n) = 2sf(s, \vec{b}) \frac{ \textstyle
i^{n-1}}{ \textstyle \sqrt{n!}} \prod_{i=1}^{n} J( s, \vec{b}; q_{i}),
\end{equation}
where, owing to unitarity,
\begin{equation}
\mid f(s, \vec{b}) \mid^{2} = e^{ \textstyle - \overline{n}(s, \vec{b})},
\end{equation}
and
\begin{equation}
\overline{n}(s, \vec{b}) = \int dq \mid J(s, \vec{b};q) \mid^{2}
\end{equation}
denotes the average number of emitted pions at a given impact
parameter $b$.
The function $ \mid J(s, \vec{b}; q) \mid^{2}$,after the integration over b,
controls the shape of the single-particle inclusive distribution.
A suitable choice of this function also
guarantees that the energy and the momentum
are conserved on the average during the collision.

The above particular factorization of $T_{n}$ was considered earlier by
R. Avin et al. [19] in connection with the construction of a unitary
model of multiparticle production.

The inclusion of isospin in this model is straightforward [3].
We observe that the factorization of $T_{n}$ is a consequence of
the pion field satisfying the equation of motion
\begin{equation}
( \Box + \mu^{2}) \vec{ \pi}( s,\vec{b}; x) = \vec{j}(s,
\vec{b};x),
\end{equation}
where $ \vec{j}$ is a classical source related to $ \vec{J}(s,
\vec{b};q )$ via the Fourier transform
\begin{equation}
\vec{J}(s, \vec{b};q) = \int d^{4}x e^{ iqx}
\vec{j}(s, \vec{b};x).
\end{equation}

The standard solution of Eq.(8) is given in terms of in- and
out-fields that are connected by the unitary
$S$ matrix $ \hat{S}( \vec{b},s)$ as follows:
\begin{equation}
\vec{ \pi}_{out} = \hat{S}^{ \dagger} \vec{ \pi}_{in}
\hat{S} = \vec{ \pi}_{in} + \vec{ \pi}_{classical},
\end{equation}
where
\begin{equation}
\vec{ \pi}_{classical} = \int d^{4}x' \Delta(x-x'; \mu) \vec{j}(s,
\vec{b};x').
\end{equation}

The $S$ matrix following from such a classical source
is still an operator in the space of pions.
Inclusion of isospin requires $ \hat{S}(s, \vec{b})$
to be also a matrix in the isospace of the leading particles.

The coherent production of isovector clusters of pions
is described by the following $S$ matrix:
\begin{equation}
\hat{S}(s, \vec{b}) = \int d^{2} \vec{e} \mid \vec{e}
\rangle D( \sum_{c} \vec{J_{c}}; s, \vec{b}) \langle \vec{e} \mid,
\end{equation}
where $ \mid \vec{e} \, \rangle $ represents the isospin-state
vector of the two-leading-particle system.
The quantity $D( \sum_{c} \vec{J_{c}};s, \vec{b})$ is the
unitary coherent-state displacement operator defined as
\begin{equation}
D( \sum_{c} \vec{J_{c}};s, \vec{b}) = exp[ \sum_{c }\int dq
\vec{J_{c}}(s, \vec{b};q) \vec{a_{c}}^{ \dagger}(q) -H.c.],
\end{equation}
where $ \vec{a_{c}}^{ \dagger}(q)$ is the creation operator of
a cluster c and the summation $ \sum_{c}$ is over all clusters.
The clusters decay independently into $ c = 1,2, \ldots $ pions
outside the region of strong interactions.

Clusters decaying into two or more pions simulate a short-range correlation
between pions and behave as $ \vec{ \rho}, \, \vec{ \rho'}, \ldots $, etc.
in isospace. They need
not be well-defined resonances. The more pions in a cluster, the
larger the correlation effect expected. If the conservation of
isospin is a global property of the colliding system, restricted only by
the relation
\begin{equation}
\vec{I} = \vec{I'} + \vec{I_{ \pi}},
\end{equation}
where $ \vec{I_{ \pi}}$ denotes the isospin of the emitted pion
cloud, then $ \vec{J_{c}}(s, \vec{b};q)$ is of the form
\begin{equation}
\vec{J_{c}}(s, \vec{b};q) = J_{c}(s, \vec{b};q) \vec{e},
\end{equation}
where $ \vec{e}$ is a fixed unit vector in isospace independent of q.
The global conservation of isospin thus introduces the long-range correlation
between the emitted pions.

If the isospin of the system of two incoming ( outgoing ) leading
particles is $II_{3}$ and $I'I'_{3}$, respectively, then the initial-state
vector of the pion field is $ \hat{S}(s, \vec{b}) \mid II_{3} \rangle,$
where $ \mid II_{3} \rangle$ is a vacuum state with no pions but with two
leading particles in the isostate characterized by $II_{3}.$ Then the
$n$-pion production amplitude is
\begin{equation}
iT_{n}(s, \vec{b};q_{1} \ldots q_{n}) = 2s \langle I'I'_{3};q_{1} \ldots
q_{n} \mid \hat{S}(s, \vec{b}) \mid II_{3} \rangle.
\end{equation}
The unnormalized probability distribution of producing $n_{+} \pi^{+}, \,
n_{ \_} \pi^{ \_},$ and $n_{0} \pi^{0}$ pions is defined as
\begin{eqnarray}
W(n_{+}n_{ \_}n_{0}, I'I'_{3},II_{3}) & \! \! = \! \! & \int d^{2}bdq_{1}dq_{2}
\ldots dq_{n} \mid \langle I'I'_{3}n_{+}n_{ \_}n_{0} \mid \hat{S}(s,
\vec{b}) \mid II_{3} \rangle \mid^{2}, \\
 \mbox{where} \hspace{2.5cm} n & \! \! = \! \! & n_{+} + n_{ \_} + n_{0}.
\nonumber
\end{eqnarray}

Assuming further that all $(I', I_{3}')$ are produced with
equal probability, we can sum over all possible isospin states of
the outgoing leading particles to obtain \\
\begin{equation}
P_{II_{3}}(n_{+}n \_ n_{0} ) = \frac{
\sum_{I'I'_{3}}W(n_{+}n\_ n_{0},I'I'_{3};II_{3})}{
\sum_{n_{+}n \_ n_{0}} \sum_{I'I'_{3}} W(n_{+} n \_
n_{0}, I'I'_{3};I I_{3} )}.
\end{equation} \\
This is  our basic relation for calculating various pion-multiplicity
distributions, pion multiplicities, and pion
correlations between definite charge combinations.
\footnote{In general, the probability $W(n_{+}n_{ \_}n_{0},I'I_{3}';II_{3})$
depends on $(I'I_{3}')$ dynamically. The final-leading-particle
production mechanism usually tends to favor the $(I',I_{3}') \approx
(I,I_{3})$ case, for example, if the final-leading particles are
nucleons or isobars. However, if the leading particles are
colliding nuclei, it is reasonable to assume almost equal
probability for various $(I',I_{3}')$ owing to the large number
of possible leading isobars in the final state.}

In order to obtain some more detailed results for multiplicity
distributions and correlations, one should have an explicit
form for the source function $J_{c}(s, \vec{b};q), c=1,2, \ldots \,.$

The results on the isospin structure are most easily
analyzed in the so-called grey-disk model in which
\begin{equation}
\overline{n}_{c}(s, \vec{b}) = \overline{n}_{c}(s) \theta(b_{o}(s)-b),
\end{equation}
where $ \overline{n}_{c}(s)$ denotes the mean number of clusters
of the type $c$, $b_{0}(s)$ is related to the total inelastic
cross section, and $ \theta$ is a step function.

Let us assume that pions are produced both directly and through isovector
clusters of the $ \rho$ type. In this case, the distribution of
neutral pions becomes
\begin{eqnarray}
P_{II_{3}}(n_{0}) & = & (I + \frac{1}{2}) \frac{(I-I_{3})!}{(I+I_{3})!}
\int_{-1}^{1}dx \mid P_{I}^{I_{3}}(x) \mid^{2} \frac{ \textstyle
\mbox{$[A(x)]$}^{n_{0}}}{ \textstyle n_{0}!}e^{ \textstyle - A(x)},
\end{eqnarray}
where $A(x) = \overline{n}_{ \pi}x^{2}+(1-x^{2}) \overline{n}_{ \rho}.$
Here $ \overline{n}_{ \pi}$ denotes the average number of directly
produced pions, and $ \overline{n}_{ \rho}$ denotes the average number
of $ \rho$-type clusters which decay into two short-range correlated
pions. The function $P_{I}^{I_{3}}(x)$ denotes the associate
Legendre polynomial.

Note that the total number of emitted pions is
\begin{equation}
\langle n \rangle = \overline{n}_{ \pi} + 2 \overline{n}_{ \rho}.
\end{equation}
The correlation $f_{2}$ between the pions,
\begin{equation}
f_{2} = \langle n(n-1) \rangle - \langle n \rangle^{2}
= 2 \overline{n}_{ \rho},
\end{equation}
is positive indicating deviation from the Poisson distribution
$( f_{2}=0 )$ owing to the emission of $ \rho$-type clusters.

In Fig. 1 we show the behavior of $P_{II_{3}}(n_{0})$ for
$ \langle n \rangle = 50$ and different combinations of $( \overline{n}_{ \pi},
\overline{n}_{ \rho}) $ when (a) $I = I_{3} = 0, $ and  ( b) $I = I_{3} = 1.$
We see that Centauro-type behavior is obtained only for pions that
are produced directly, i.e., for $ \overline{n}_{ \pi} \neq 0$ and
$ \overline{n}_{ \rho}=0.$

If $ R = \frac{ \textstyle n_{0}}{ \textstyle \langle n \rangle}$
denotes the fraction of neutral pions, then it is easy to see that
in the limit $ \langle n \rangle \rightarrow \infty$ with $R$
fixed, the probability distribution $ \langle n \rangle P_{II_{3}}(
n_{0})$ scales to the limiting behavior:
\begin{equation}
\langle n \rangle P_{II_{3}}(n_{0}) \rightarrow P_{II_{3}}(R) =
(I + \frac{1}{2}) \frac{(I-I_{3})!}{(I+I_{3})!} \mid P_{I}^{I_{3}}(
x_{R}) \mid^{2} \mbox{$[(1-3 \gamma_{ \rho})(R- \gamma_{ \rho})]$}^{
- \frac{1}{2}},
\end{equation}
where $ \gamma_{ \rho}= \frac{ \textstyle \overline{n}_{ \rho}}{
\textstyle \langle n \rangle } = \frac{ \textstyle
\overline{n}_{ \rho}}{ \textstyle ( \overline{n}_{ \pi}+
2 \overline{n}_{ \rho})}$ and
\begin{equation}
x_{R}^{2} = \frac{ \textstyle R - \gamma_{ \rho}}{ \textstyle
1 - 3 \gamma_{ \rho}} \; , \; \; \; 0 \leq \gamma_{ \rho} < \frac{1}{3}.
\end{equation}
This limiting probability distribution is different from the usual
Gaussian random distribution for which one expects peaking at
$R = \frac{1}{3}$ as $ \langle n \rangle \rightarrow \infty.$

Instead, we find peaking at $R = \gamma_{ \rho}$ although
\begin{eqnarray}
\langle n_{0} \rangle_{II} & = & \langle n \rangle - \langle
n_{ch} \rangle_{II} \nonumber \\
& = & \frac{ \textstyle 1}{ \textstyle 2I + 3} \mbox{$[
\overline{n}_{ \pi} + 2(I+1) \overline{n}_{ \rho}]$} \\
\mbox{and} \; \; \; \;  \langle n_{0} \rangle & = & \frac{1}{2}
\langle n_{ch} \rangle = \frac{1}{3} \langle n \rangle,
\; \; \; \mbox{for either } \; I=0 \; \mbox{or} \;
\overline{n}_{\pi} = \overline{n}_{ \rho}. \nonumber
\end{eqnarray}
If $ \gamma_{ \rho} = \frac{1}{3},$ that is if $ \overline{n}_{ \pi}
= \overline{n}_{ \rho},$ the probability $P_{II_{3}}(n_{0})$ becomes a pure
Poisson distribution and is independent of $II_{3}.$

In Fig. 2 we show the average number of neutral pions
$( \langle n_{0} \rangle_{n_{ \_}}),$ as a function of the number of
negative pions $(n_{ \_})$ produced for different pairs of
$( \overline{n}_{ \pi}, \overline{n}_{ \rho})$ and $I=I_{3}=1.$ We note
that $ \langle n_{0} \rangle_{ n_{ \_}}$ decreases with $n_{ \_}$ only in
the region where $ \overline{n}_{ \pi} \gg \overline{n}_{ \rho }$
and $ \langle n \rangle \gg n_{ \_},$ indicating again that the Centauro-type
effect cannot be expected if a significant number of $ \rho$-type
clusters of pions is produced. Similarly we find that
the neutral-pion two-particle correlation parameters
$f_{2}^{0a} = \langle n_{0} n_{a} \rangle - \langle n_{0}
\rangle \langle n_{a} \rangle - \langle n_{0} \rangle
\delta_{0a}$ for $a = 0$ and $-$ are related as
\begin{equation}
f_{2,II}^{0-} + \frac{1}{2} f_{2,II}^{00} = \overline{n}_{
\rho} \frac{ \textstyle I+1}{ \textstyle 2I +3},
\end{equation}
where
\begin{equation}
f_{2,II}^{00} = \frac{ \textstyle 4(I+1)}{ \textstyle
(2I+3)^{2}(2I+5)}( \overline{n}_{ \pi} - \overline{n}_{
\rho})^{2},
\end{equation}
showing that $f_{2,II}^{0-}$ is negative if $ \overline{n}_{\rho}$
is small. We should also mention earlier works on
neutral-pion correlation effects [20].

In summary, the results of the present analysis show that the
emission of isovector clusters of pions in the framework of a unitary
eikonal model with the global conservation of isospin strongly suppreses
the Centauro-type effect for pions. This might suggest that the Centauro-type
effect, if any, will probably appear only in very limited regions
of phase space where isovector clusters should be missing. How it is
possible dynamically is not yet clear [10--13]. Similar discussion
about emission of isoscalar clusters of pions is presented in [15].

{ \large \bf Acknowledgment }

This work was supported by the Ministry of Science of
Croatia under Contract No. 1 - 03 - 212.

\newpage

{\bf Figure captions :}

Fig. 1. Multiplicity distributions $P_{II_{3}}(n_{0})$ of neutral pions
for $ \langle n \rangle = 50$ when the total isospin of the
incoming-leading-particle system
is $(a) \; \; I = I_{3} = 0$ and $(b) \; \; I = I_{3} = 1.$ The curve s
represent different combinations of $( \overline{n}_{ \pi}, \overline{n}_{
\rho}),$ the average number of directly produced pions, and the average number
of $ \rho$-type clusters, respectively.

Fig. 2. The average number of neutral pions as a function of the number
of negative pions produced for $I = I_{3} = 1.$ The curves represent
different combinations of $( \overline{n}_{ \pi}, \overline{n}_{ \rho}).$

\end{document}